\documentstyle[11pt]{article}
\newtheorem{theorem}{Theorem}

\newcommand{\edfn}{\stackrel{\triangle}{=}}

\newcommand{\lea}{\stackrel{\mbox{(a)}}{\le}}
\newcommand{\leb}{\stackrel{\mbox{(b)}}{\le}}
\newcommand{\lec}{\stackrel{\mbox{(c)}}{\le}}
\newcommand{\led}{\stackrel{\mbox{(d)}}{\le}}
\newcommand{\lee}{\stackrel{\mbox{(e)}}{\le}}
\newcommand{\lef}{\stackrel{\mbox{(f)}}{\le}}

\newcommand {\bx} {\mbox{\boldmath $x$}}
\newcommand {\by} {\mbox{\boldmath $y$}}

\newcommand {\bE} {\mbox{\boldmath $E$}}

\newcommand {\bX} {\mbox{\boldmath $X$}}
\newcommand {\bY} {\mbox{\boldmath $Y$}}

\newcommand{\calA}{{\cal A}}

\newcommand{\calC}{{\cal C}}

\newcommand{\calE}{{\cal E}}

\newcommand{\calR}{{\cal R}}

\newcommand{\calX}{{\cal X}}
\newcommand{\calY}{{\cal Y}}

\begin{document}
\thispagestyle{empty}
\setcounter{page}{1}
\setlength{\baselineskip}{1.5\baselineskip}
\title{Universal Decoding With an Erasure Option
\thanks{This research was supported by the Israel Science Foundation, grant no.\ 223/05.}}
\author{Neri Merhav\thanks{Department of Electrical Engineering, 
Technion -- Israel Institute of Technology, Haifa 32000, Israel.
E--mail: [merhav@ee.technion.ac.il].}
\and
Meir Feder\thanks{Department of Electrical Engineering -- Systems, Tel Aviv University,
Tel Aviv 69978, Israel. E-mail: [meir@eng.tau.ac.il].}} 
\maketitle

\begin{abstract}
Motivated by applications of rateless coding, decision feedback, and ARQ, we 
study the problem of universal decoding for unknown channels, in the presence
of an erasure option. 
Specifically, we harness the competitive minimax methodology developed
in earlier studies, in order to derive a universal version of Forney's classical
erasure/list decoder, which in the erasure case, optimally trades off between the probability
of erasure and the probability of undetected error.
The proposed universal erasure decoder guarantees
universal achievability of a certain fraction $\xi$ of the 
optimum error exponents of these probabilities
(in a sense to be made precise in the sequel).
A single--letter expression for $\xi$, which depends solely on the
coding rate and the threshold, is provided.
The example of the binary symmetric channel
is studied in full detail, and some conclusions are drawn.\\

\end{abstract}

\clearpage
\section{Introduction}

When communicating across an unknown channel,
classical channel coding at any fixed rate, however small, is inherently problematic since
this fixed rate might be larger than the unknown capacity of the underlying channel.
It makes sense then to try to adapt the coding rate to the channel conditions,
which can be learned on--line at the transmitter whenever a feedback 
link, from the receiver to the transmitter, is available. 

One of the recent promising 
approaches to this end is rateless coding (see, e.g., \cite{DFK}, \cite{DFK04},
\cite{EWT05}, \cite{JN06}, \cite{Shulman03}, \cite{SF00},
\cite{TT04}, and references 
therein). 
According to this approach,
there is fixed number of messages $M$, each one being represented by a codeword
of unlimited length, in principle. After each transmitted symbol of the message
selected, the decoder examines whether it can make a decision, namely, decode
the message, with ``reasonably good confidence,'' 
or alternatively, to request, via the feedback link,
an additional symbol to be transmitted, before arriving at a decision. 
Upon receiving the new channel output, again, the receiver either makes a decision, or
requests another symbol from the transmitter, and so on.\footnote{Alternatively, the
receiver can use the feedback link only to notify the transmitter when it reached
a decision regarding the current message (and keep silent at all other times).
In network situations, this would not load the
network much as it is done only once per each message.}
The coding rate, in such
a scenario, is defined as $\log M$ divided by the expected number of symbols transmitted
before the decoder finally commits to a decision.
Clearly, at every time instant, the receiver of a rateless communication system operates
just like an {\it erasure decoder} \cite{Forney68},\footnote{See also \cite{Viterbi69},
\cite{ACZ96}, \cite{HT97}, \cite{Hashimoto99} and referecnes therein for later studies.}
which partitions the space of
channel output vectors into $(M+1)$ regions, $M$ for each one of the possible
messages, and an additional region for ``erasure,'' i.e., ``no decision,'' which
in the rateless regime, is used for requesting additional information from the transmitter.
Keeping the erasure probability small is then motivated by the desire to keep the expected
transmission time, for each message, small. Although these two criteria are not completely equivalent, they
are nonetheless strongly related.

This observation, as well as techniques such as ARQ and decision feedback,
motivate us to study the problem of universal decoding with an
erasure option, for the class of discrete memoryless channels (DMC's)
indexed by an unknown parameter vector $\theta$ (e.g., the set of channel transition probabilities).
Specifically, we harness the competitive minimax methodology proposed
in \cite{FM02}, in order to derive a universal version of Forney's classical
erasure/list decoder. For a given DMC with parameter $\theta$, a given coding rate $R$,
and a given threshold parameter $T$ (all to be formally defined later),
Forney's erasure/list decoder optimally 
trades off between the exponent $E_1(R,T,\theta)$ of the probability
of the erasure event, $\calE_1$, and the exponent, $E_2(R,T,\theta)=E_1(R,T,\theta)+T$,
of the probability of undetected error event, $\calE_2$, 
in the random coding regime.

The universal erasure decoder, proposed in this paper, guarantees
universal achievability of an erasure exponent, $\hat{E}_1(R,T,\theta)$,
which is at least as large as $\xi\cdot E_1(R,T,\theta)$ for all $\theta$, for some
constant $\xi\in(0,1]$, that is independent of $\theta$ (but does depend on $R$ and $T$),
and at the same time,
an undetected error exponent $\hat{E}_2(R,T,\theta)\ge \xi\cdot E_1(R,T,\theta)+T$
for all $\theta$ (in the random coding sense). At the very least this guarantees
that whenever the probabilities of $\calE_1$ and $\calE_2$ decay exponentially
for a known channel, so they do even when the channel is unknown, using the proposed universal
decoder. The question is, of course: what is the largest value of $\xi$ for which the
above statement holds? We answer this question by deriving a
single--letter expression for a lower bound to the largest value
of $\xi$, denoted henceforth by $\xi^*(R,T)$, that is guaraneteed to be attainable by this
decoder. It is conjectured that $\xi^*(R,T)$
reflects the best fraction of $E_1(R,T,\theta)$ (and of $E_2(R,T,\theta)$ in the
above sense) that any decoder that is unaware of $\theta$ can uniformly achieve.
Explicit results, including numerical values of $\xi^*(R,T)$,
are derived for the example of the binary symmetric channel (BSC), parameterized
by the crossover probability $\theta$,
and some conclusions are drawn.

The outline of the paper is as follows. In Section 2, we establish the notation conventions
and we briefly review some known results about erasure decoding. 
In Section 3, we formulate the problem of
universal decoding with erasures. In Section 4, we present the proposed universal erasure decoder
and prove its asymptotic optimality in the competitive minimax sense.
In Section 5, we present the main results concering the performance
of the proposed universal decoder. Section 6 is devoted to the example
of the BSC. Finally, in Section 7, we summarize our conclusions.

\section{Notation and Preliminaries}

Throughout this paper, scalar random variables (RV's) will be denoted by capital
letters, their sample values will be denoted by
the respective lower case letters, and their alphabets will be denoted
by the respective calligraphic letters.
A similar convention will apply to
random vectors of dimension $n$ and their sample values,
which will be denoted with same symbols in the bold face font.
The set of all $n$--vectors with components taking values in a certain alphabet,
will be denoted as the same alphabet superscripted by $n$.
Thus, for example, a random vector $\bX=(X_1,\ldots,X_n)$ may assume
a specific vector value $\bx=(x_1,\ldots,x_n)\in\calX^n$ as each
component takes values in $\calX$.
Channels will be denoted generically by the letter $P$, or $P_\theta$, when
we wish to emphasize that the channel is indexed or parametrized by 
a certain scalar or vector $\theta$, taking on values in some set $\Theta$.
Information theoretic quantities like entropies and conditional entropies,
will be denoted following the usual conventions
of the information theory literature, e.g., $H(X)$, $H(X|Y)$, and so on.
The cardinality of a finite set $\calA$ will be denoted by $|\calA|$.

Consider a discrete memoryless channel (DMC) with a finite 
input alphabet $\calX$,
finite output alphabet $\calY$, and single--letter transition probabilities
$\{P(y|x),~x\in\calX,~y\in\calY\}$. As the channel is fed by an input vector
$\bx\in\calX^n$, it generates an output vector 
$\by\in\calY^n$ according to the conditional probability distribution
\begin{equation}
P(\by|\bx)=\prod_{i=1}^n P(y_i|x_i).
\end{equation}
A rate--$R$ block code of length $n$ consists of $M=e^{nR}$ $n$--vectors
$\bx_m\in\calX^n$, $m=1,2,\ldots,M$, which represent $M$ different messages.
We will assume that all possible messages are
a--priori equiprobable, i.e., $P(m)=1/M$ for all $m=1,2,\ldots,M$.

A decoder with an erasure option is a partition of $\calY^n$ into $(M+1)$ regions,
$\calR_0,\calR_1,\ldots,\calR_M$. Such a decoder works as follows: If $\by$ falls
into $\calR_m$, $m=1,2,\ldots,M$, then a decision is made in favor of message number $m$.
If $\by\in\calR_0$, no decision is made and an erasure is declared. We will refer to
$\calR_0$ as the {\it erasure event}.

Given a code $\calC=\{\bx_1,\ldots,\bx_M\}$ 
and a decoder$ \calR=(\calR_0,\calR_1,\ldots,\calR_m)$, let us now define
two additional undesired events. 
The event $\calE_1$ is the event of not making the right decision. This
event is the disjoint union of the erasure event and the event
$\calE_2$, which is the
{\it undetected error} event, namely, the event of making the wrong decision.
The probabilities of all three events are defined as follows:
\begin{eqnarray}
\mbox{Pr}\{\calE_1\}&=&\frac{1}{M}\sum_{m=1}^M\sum_{\by\in\calR_m^c}P(\by|\bx_m)\\
\mbox{Pr}\{\calE_2\}&=&\frac{1}{M}\sum_{m=1}^M\sum_{\by\in\calR_m}\sum_{m'\ne m}P(\by|\bx_{m'})\\
\mbox{Pr}\{\calR_0\}&=&\mbox{Pr}\{\calE_1\}-\mbox{Pr}\{\calE_2\}.
\end{eqnarray}
Forney \cite{Forney68} assumes that the DMC is known to the decoder, and
shows, using the Neyman--Pearson methodology, 
that the best
tradeoff between $\mbox{Pr}\{\calE_1\}$ and $\mbox{Pr}\{\calE_2\}$ (or, equivalently,
between $\mbox{Pr}\{\calR_0\}$ and $\mbox{Pr}\{\calE_2\}$) is attained by the decoder
$\calR^*=(\calR_0^*,\calR_1^*,\ldots,\calR_M^*)$ defined by
\begin{eqnarray}
\calR_m^*&=&\left\{\by:~\frac{P(\by|\bx_m)}{\sum_{m'\ne m} 
P(\by|\bx_{m'})}\ge e^{nT}\right\},~~m=1,2,\ldots,M\nonumber\\
\calR_0^*&=&\bigcap_{m=1}^M (\calR_m^*)^c,
\end{eqnarray}
where $(\calR_m^*)^c$ is the complement of $\calR_m^*$, and
where $T\ge 0$ is a parameter, henceforth referred to as the {\it threshold},
which controls the balance between the probabilities of $\calE_1$ and $\calE_2$.

Forney devotes the remaining part of his paper \cite{Forney68} 
to derive lower bounds
to the random coding
exponents (associated with $\calR^*$), 
$E_1(R,T)$ and $E_2(R,T)$, of $\overline{\mbox{Pr}}\{\calE_1\}$ and $\overline{\mbox{Pr}}\{\calE_2\}$,
the average\footnote{Here, ``average'' means w.r.t.\ the ensemble of randomly selected codes.}
probabilities of $\calE_1$ and $\calE_2$, respectively,
and to investigate their properties. Specifically, Forney shows, among other
things, that for the ensemble of randomly chosen codes, where each codeword is chosen
independently under an i.i.d.\ distribution $Q^n(\bx)=\prod_{i=1}^nQ(x_i)$,
\begin{equation}
E_1(R,T)=\max_{0\le s\le \rho\le 1}\max_Q [E_0(s,\rho,Q)-\rho R-sT]
\end{equation}
where 
\begin{equation}
E_0(s,\rho,Q)=-\ln\left[\sum_{y\in\calY}\left(\sum_{x\in\calX}Q(x)P^{1-s}(y|x)\right)\cdot\right.
\left.\left(\sum_{x'\in\calX}Q(x')P^{s/\rho}(y|x')\right)^\rho\right],
\end{equation}
and
\begin{equation}
E_2(R,T)=E_1(R,T)+T.
\end{equation}
A simple observation that we will need, before passing to the case of an unknown channel,
is that the same decision rule $\calR^*$ would be obtained if
rather than adopting the Neyman--Pearson approach, one would consider a Lagrange function,
\begin{equation}
\Gamma(\calC,\calR)\edfn\mbox{Pr}\{\calE_2\}+e^{-nT}\mbox{Pr}\{\calE_1\},
\end{equation}
for a given code $\calC=\{\bx_1,\ldots,\bx_M\}$ and a given threshold $T$,
as the figure of merit, and seek a decoder $\calR$ that minimizes it.
To see that this is equivalent, let us
rewrite $\Gamma(\calC,\calR)$ as follows:
\begin{equation}
\Gamma(\calC,\calR)=\frac{1}{M}\sum_{m=1}^M\left[\sum_{\by\in\calR_m}\sum_{m'\ne m}P(\by|\bx_{m'})+
\sum_{\by\in\calR_m^c}e^{-nT}P(\by|\bx_m)\right],
\end{equation}
and it is now clear that for each $m$, the bracketed expression
(which has the form of weighted error of a binary hypothesis testing problem)
is minimized by $\calR_m^*$
as defined above. Since this decision rule is identical to Forney's
one, it is easy to see that the resulting exponential decay of 
the ensemble average
$$\bE\{\Gamma(\calC,\calR^*)\}=
\overline{\mbox{Pr}}\{\calE_2\}+e^{-nT}\overline{\mbox{Pr}}\{\calE_1\}$$
is $E_2(R,T)$, as $\overline{\mbox{Pr}}\{\calE_1\}$ decays according to 
$e^{-nE_1(R,T)}$,
$\overline{\mbox{Pr}}\{\calE_2\}$ decays according to 
$e^{-nE_2(R,T)}$, and $E_2(R,T)=E_1(R,T)+T$, as mentioned earlier.
This Largrangian approach will be more convenient to work with, when we next
move on to the case of an unknown DMC, because it allows as to work with one figure
of merit instead of a trade--off between two.

\section{Unknown Channel -- Problem Description}

We now move on to the case of an unknown channel. While our techniques can be
applied to quite general classes of channels, here, 
for the sake of concreteness and conceptual simplicity, and following in \cite{Forney68},
we confine attention to DMC's.
Consider then a family of DMC's
$\{P_\theta(y|x),~x\in\calX,~y\in\calY,~\theta\in\Theta\}$, 
where $\theta$ is the parameter, or the index of the channel in the class,
taking values in some set $\Theta$. For example, $\theta$ may be a positive integer,
denoting the index of the channel within a finite 
or a countable index set. As another example,
$\theta$ may simply represent
the set of all $|\calX|\cdot(|\calY|-1)$
single--letter transition probabilties 
that define the DMC, and if there are some symmetries
(like in the BSC), these reduce the dimensionality of $\theta$.
The basic questions are now the following: 
\begin{enumerate}
\item How to devise a good erasure decoder when the
underlying channel is known to belong to the 
class $\{P_\theta(y|x),~x\in\calX,~y\in\calY,~\theta\in\Theta\}$, but
$\theta$ is unknown? 
\item What are the resulting error exponents of $\calE_1$
and $\calE_2$ and how do they compare to Forney's exponents for known $\theta$?
\end{enumerate}

In the quest for universal schemes for decoding with an erasure option,
two difficulties are encountered in light of \cite{Forney68}.
The first difficulty is that here we have two
figure of merits, the probabilities of $\calE_1$ and $\calE_2$. But this difficulty
can be alleviated by adopting the Lagrangian approach, described at the end of the
previous section.
The second difficulty is somewhat deeper: 
Classical derivations of universal 
decoding rules for ordinary decoding (without erasures)
over the class of DMC's, 
like the maximum mutual information (MMI) decoder \cite{CK81} and its variants, were
based on ideas that are deeply rooted in considerations of joint typicality
between the channel output $\by$ and each 
hypothesized codeword $\bx_m$. These considerations
were easy to apply in ordinary decoding, where the score function 
(or, the ``metric'') associated with
the optimum maximum likelihood (ML) decoding, 
$\log P_\theta(\by|\bx_m)$, involves only {\it one} codeword at a time,
and that this function
depends on $\bx_m$ and $\by$ only via their 
joint empirical distribution, or, in other words, their joint type. 
Moreover, in the case of decoding without erasures,
given the true transmitted codeword $\bx_m$ and the 
resulting channel output $\by$,
the scores associated with all other randomly chosen codewords, are independent of each other, a
fact that facilitates the analysis to a great extent.
This is very different from
the situation in erasure decoding, where Forney's optimum score function for each codeword,
$$\frac{P_\theta(\by|\bx_m)}{\sum_{m'\ne m}P_\theta(\by|\bx_{m'})},$$
depends on
{\it all} codewords at the same time. Consequently, in a random coding analysis, it is
rather complicated to apply joint typicality considerations, or to
analyze the statistical behavior of this expression, let alone the statistical
dependency between the score functions associated with the various codewords.

This difficulty is avoided if the competitive minimax methodology, proposed
and developed in \cite{FM02}, is applied.
Specifically, let $\Gamma_\theta(\calC,\calR)$ denote the
above defined Lagrangian, where we now emphasize the dependence on
the index of the channel, $\theta$.
Let us also define
$\bar{\Gamma}_\theta^*=\bE\{\min_{\calR}\Gamma_\theta(\calC,\calR)\}$, i.e., the
ensemble average of the minimum of
the above Lagrangian (achieved by Forney's optimum decision rule) w.r.t.\
the channel $\{P_\theta(y|x)\}$ for a given $\theta$. Note that the exponential order
of $\bar{\Gamma}_\theta^*$ is $e^{-n[E_1(R,T,\theta)+T]}=e^{-nE_2(R,T,\theta)}$, where
$E_1(R,T,\theta)$ and
$E_2(R,T,\theta)$ are
the new notations for $E_1(R,T)$ and $E_2(R,T)$, respectively, with the dependence on the
channel index $\theta$, made explicit. 
In principle, we would have been interested in a decision rule $\calR$ that achieves
\begin{equation}
\min_{\calR}\max_{\theta\in\Theta}\frac{\Gamma_\theta(\calC,\calR)}
{\bar{\Gamma}_\theta^*},
\end{equation}
or, equivalently,
\begin{equation}
\min_{\calR}\max_{\theta\in\Theta}\frac{\Gamma_\theta(\calC,\calR)}
{e^{-n[E_1(R,T,\theta)+T]}},
\end{equation}
but as is discussed in \cite{FM02} (in the analogous context of ordinary decoding, without erasures),
such an ambitious minimax criterion of competing with the
optimum performance may be too optimisitic. 
A better approach would be to
compete with a similar expression of the exponential behavior, but where the term
$E_1(R,T,\theta)$ is being multiplied by a constant $\xi\in(0,1]$, which we would like 
to choose as large as possible. In other words,
we are interested in the 
competitive minimax criterion
\begin{equation}
\label{kn}
K_n(\calC)\edfn\min_{\calR}\max_{\theta\in\Theta}\frac{\Gamma_\theta(\calC,\calR)}
{e^{-n[\xi E_1(R,T,\theta)+T]}}.
\end{equation}
Similarly as in \cite{FM02}, we
wish to find the largest value of $\xi$ such that 
the ensemble average $\bar{K}_n\edfn\bE\{K_n(\calC)\}$ would not grow exponentially fast,
i.e.,
\begin{equation}
\limsup_{n\to\infty}\frac{1}{n}\log \bar{K}_n \le 0.
\end{equation}
The rationale behind this is the following: 
If $\bar{K}_n$ is sub--exponential in $n$, for some $\xi$, then
this guarantees that there exists a universal erasure decoder,
say $\hat{\calR}$, such that for {\it every}
$\theta\in\Theta$,
the exponential order of $\bE\{\Gamma_\theta(\calC,\hat{\calR})\}$ 
is no worse than
$e^{-n[\xi E_1(R,T,\theta)+T]}$. This, in turn, implies
that {\it both} terms of $\Gamma_\theta(\calC,\hat{\calR})$ 
decay at least as  $e^{-n[\xi E_1(R,T,\theta)+T]}$,
which means that for the decoder $\hat{\calR}$, the exponent of $\overline{\mbox{Pr}}\{\calE_1\}$
is at least $\xi\cdot E_1(R,T,\theta)$ and the exponent of $\overline{\mbox{Pr}}\{\calE_2\}$
is at least $\xi\cdot E_1(R,T,\theta)+T$, 
both for every $\theta\in\Theta$.
Thus, the difference between the two (guaranteed) 
exponents remains $T$ as before (as the weight of the
term $\overline{\mbox{Pr}}\{\calE_1\}$ in $\Gamma(\calR,\calC)$ is $e^{-nT}$), 
but the other term, $E_1(R,T,\theta)$, is now scaled by a factor of $\xi$.

The remaining parts of this paper focus on deriving a universal
decoding rule that asymptotically achieves $\bar{K}_n(\calC)$ for a given $\xi$,
and on analyzing its performance, i.e., finding the maximum value of $\xi$
such that $\bar{K}_n$ still grows sub--exponentially rapidly.

\section{Derivation of a Universal Erasure Decoder}

For a given $\xi\in(0,1]$, let us define
\begin{equation}
f(\bx_m,\by)\edfn\max_{\theta\in\Theta}\left\{e^{n[\xi E_1(R,T,\theta)+T]}P_\theta(\by|\bx_m)\right\}
\end{equation}
and consider the decoder
\begin{eqnarray}
\hat{\calR}_m&=&\left\{\by:~\frac{f(\bx_m,\by)}{\sum_{m'\ne m} 
f(\bx_{m'},\by)}\ge e^{nT}\right\},~~m=1,2,\ldots,M\nonumber\\
\hat{\calR}_0&=&\bigcap_{m=1}^M \hat{\calR}_m^c.
\end{eqnarray}
Denoting
\begin{equation}
\label{knr}
K_n(\calC,\calR)=
\max_{\theta\in\Theta}\frac{\Gamma_\theta(\calC,\calR)}
{e^{-n[\xi E_1(R,T,\theta)+T]}},
\end{equation}
for a given encoder $\calC=\{\bx_1,\ldots,\bx_M\}$ and decoder $\calR$,
our first main result estabilishes the asymptotic optimality of $\hat{\calR}$
in the competitive minimax sense, namely, that $K_n(\calC,\hat{\calR})$ is within
a sub--exponential factor as small as $K_n(\calC)=\min_{\calR}K_n(\calC,\calR)\}$,
and therefore, $\bE\{K_n(\calC,\hat{\calR})\}$ is within the same sub--exponential factor
as small as $\bar{K}_n=\bE\{K_n(\calC)\}$.
\begin{theorem}
For every code $\calC$,
\begin{equation}
K_n(\calC,\hat{\calR})\le (n+1)^{|\calX|\cdot|\calY|-1}K_n(\calC).
\end{equation}
\end{theorem}

\noindent
{\it Comment:}
Note that the summation $\sum_{m'\ne m}f(\bx_{m'},\by)$ might pose some
numerical challenges since it is a summation of many terms within a potentially
large range of order of magnitudes. An asymptotically equivalent version of $\hat{\calR}$,
that avoids such summations altogether, 
is the following. Let $M(\alpha)$ be the number of
$\{\bx_{m'}\}$ for which $f(\bx_{m'},\by)=\alpha$. Since $f(\bx_{m'},\by)$
depends on $(\bx_{m'},\by)$ only via their joint empirical distribution
(see the proof of Theorem 1, next), then the number of possible values of $\alpha$ is
at most polynomial in $n$.
Then, $\sum_{m'\ne m}f(\bx_{m'},\by)$
can be replaced by $\max_\alpha [\alpha\cdot M(\alpha)]$,
without affecting the asymptotic optimality.

\noindent
{\it Proof.} The proof technique is similar to that of \cite{FM02}.
As $\bx$ and $\by$ exhaust their spaces, $\calX^n$ and $\calY^n$, let
$\Theta_n$ denote set of values of $\theta$ that achieve 
$\{f(\bx,\by),~\bx\in\calX^n,~\by\in\calY^n\}$.
Observe that for every $\theta$, the expression
$[e^{n[\xi E_1(R,T,\theta)+T]}P_\theta(\by|\bx)]$ depends on $(\bx,\by)$
only via their joint empirical distribution (or, the joint type). Consequently,
the value of $\theta$ that achieves $f(\bx,\by)$ also depends on $(\bx,\by)$
only via their joint empirical distribution. Since the number of joint 
empirical distributions of $(\bx,\by)$ never exceeds 
$(n+1)^{|\calX|\cdot|\calY|-1}$ (see \cite{CK81}),
then obviously
\begin{equation}
\label{Thetan}
|\Theta_n|\le (n+1)^{|\calX|\cdot|\calY|-1}
\end{equation}
as well.
Now, for every encoder $\calC$ and decoder $\calR$,
\begin{eqnarray}
K_n(\calC,\calR)&=&\max_{\theta\in\Theta}\frac{\Gamma_\theta(\calC,\calR)}
{e^{-n[\xi E_1(R,T,\theta)+T]}}\nonumber\\
&=&\max_{\theta\in\Theta}\frac{1}{M}\sum_{m=1}^M\left[\sum_{\by\in\calR_m}\sum_{m'\ne m}
\frac{P_\theta(\by|\bx_{m'})}{e^{-n[\xi E_1(R,T,\theta)+T]}}+
e^{-nT}\sum_{\by\in\calR_m^c}\frac{P_\theta(\by|\bx_m)}{e^{-n[\xi E_1(R,T,\theta)+T]}}\right]
\nonumber\\
&\le&\frac{1}{M}\sum_{m=1}^M\left[\sum_{\by\in\calR_m}\sum_{m'\ne m}\max_{\theta\in\Theta}
\frac{P_\theta(\by|\bx_{m'})}{e^{-n[\xi E_1(R,T,\theta)+T]}}+
e^{-nT}\sum_{\by\in\calR_m^c}
\max_{\theta\in\Theta}\frac{P_\theta(\by|\bx_m)}{e^{-n[\xi E_1(R,T,\theta)+T]}}\right]\nonumber\\
&=&\frac{1}{M}\sum_{m=1}^M\left[\sum_{\by\in\calR_m}\sum_{m'\ne m}
f(\bx_{m'},\by)+
e^{-nT}\sum_{\by\in\calR_m^c}f(\bx_m,\by)\right]\nonumber\\
&\edfn&\hat{K}_n(\calC,\calR)\nonumber\\
&=&\frac{1}{M}\sum_{m=1}^M\left[\sum_{\by\in\calR_m}\sum_{m'\ne m}\max_{\theta\in\Theta_n}
\frac{P_\theta(\by|\bx_{m'})}{e^{-n[\xi E_1(R,T,\theta)+T]}}+
e^{-nT}\sum_{\by\in\calR_m^c}
\max_{\theta\in\Theta_n}\frac{P_\theta(\by|\bx_m)}{e^{-n[\xi E_1(R,T,\theta)+T]}}\right]\nonumber\\
&\le&\frac{1}{M}\sum_{m=1}^M\left[\sum_{\by\in\calR_m}\sum_{m'\ne m}\left(\sum_{\theta\in\Theta_n}
\frac{P_\theta(\by|\bx_{m'})}{e^{-n[\xi E_1(R,T,\theta)+T]}}\right)+\right.\nonumber\\
& &\left. e^{-nT}\sum_{\by\in\calR_m^c}\left(
\sum_{\theta\in\Theta_n}\frac{P_\theta(\by|\bx_m)}{e^{-n[\xi E_1(R,T,\theta)+T]}}
\right)\right]\nonumber\\
&=&\sum_{\theta\in\Theta_n}\frac{1}{M}\sum_{m=1}^M\left[\sum_{\by\in\calR_m}\sum_{m'\ne m}
\frac{P_\theta(\by|\bx_{m'})}{e^{-n[\xi E_1(R,T,\theta)+T]}}+\right. \nonumber\\
& &\left. e^{-nT}\sum_{\by\in\calR_m^c}
\frac{P_\theta(\by|\bx_m)}{e^{-n[\xi E_1(R,T,\theta)+T]}}
\right]\nonumber\\
&\le&|\Theta_n|\cdot
\max_{\theta\in\Theta_n}\frac{1}{M}\sum_{m=1}^M\left[\sum_{\by\in\calR_m}\sum_{m'\ne m}
\frac{P_\theta(\by|\bx_{m'})}{e^{-n[\xi E_1(R,T,\theta)+T]}}+\right. \nonumber\\
& &\left. e^{-nT}\sum_{\by\in\calR_m^c}
\frac{P_\theta(\by|\bx_m)}{e^{-n[\xi E_1(R,T,\theta)+T]}}
\right]\nonumber\\
&\le&(n+1)^{|\calX|\cdot|\calY|-1}\cdot
\max_{\theta\in\Theta}\frac{1}{M}\sum_{m=1}^M\left[\sum_{\by\in\calR_m}\sum_{m'\ne m}
\frac{P_\theta(\by|\bx_{m'})}{e^{-n[\xi E_1(R,T,\theta)+T]}}+\right. \nonumber\\
& &\left. e^{-nT}\sum_{\by\in\calR_m^c}
\frac{P_\theta(\by|\bx_m)}{e^{-n[\xi E_1(R,T,\theta)+T]}}
\right]\nonumber\\
&\le&(n+1)^{|\calX|\cdot|\calY|-1}\cdot K_n(\calC,\calR).
\label{longchain}
\end{eqnarray}
Thus, we have defined $\hat{K}_n(\calC,\calR)$ and sandwiched it between
$K_n(\calC,\calR)$ and $(n+1)^{|\calX|\cdot|\calY|-1}\cdot K_n(\calR,\calC)$ uniformly
for every $\calC$ and $\calR$. 
Now, obviously, $\hat{\calR}$ minimizes $\hat{K}_n(\calC,\calR)$, and so,
for every $\calR$,
\begin{equation}
K_n(\calC,\hat{\calR})\le \hat{K}_n(\calC,\hat{\calR})\le \hat{K}_n(\calC,\calR)\le 
(n+1)^{|\calX|\cdot|\calY|-1}\cdot K_n(\calC,\calR),
\end{equation}
where the first and the third inequalites were just proved in the chain of inequalities 
(\ref{longchain}), and the second inequality follows from the optimality of $\hat{\calR}$
w.r.t.\ $\hat{K}_n(\calC,\calR)$. Since we have shown that
$$K_n(\calC,\hat{\calR})\le (n+1)^{|\calX|\cdot|\calY|-1}\cdot K_n(\calC,\calR)$$
for every $\calR$,
we can now minimize the r.h.s.\ w.r.t.\ $\calR$ and the assertion of Theorem 1 is obtained.
This completes the proof of Theorem 1.

\section{Performance}

In this section, we present an upper bound to $\bar{K}_n$
from which we derive a lower bound to $\xi^*$, the largest value of $\xi$
for which $\bar{K}_n$ is sub--exponential in $n$. 

Given a distribution $P_y$ on $\calY$, a positive real $\lambda$, and a value of $\theta$, let
\begin{equation}
\label{F}
F(P_y,\lambda,\theta)\edfn\ln|\calX|-\max_{P_{x|y}}[H(X|Y)+\lambda \bE\ln P_\theta(Y|X)],
\end{equation}
where
$\bE\{\cdot\}$ is the expectation and $H(X|Y)$ is the conditional entropy w.r.t.\ a generic
joint distribution $P_{xy}(x,y)=P_y(y)P_{x|y}(x|y)$ of the RV's $(X,Y)$.
Next, for a pair $(\theta,\tilde{\theta})\in\Theta^2$, and for two real numbers
$s$ and $\rho$, $0\le s\le\rho\le 1$, define:
\begin{equation}
E(\theta,\tilde{\theta},\rho,s)=\min_{P_y}[
F(P_y,1-s,\theta)+
\rho F(P_y,s/\rho,\tilde{\theta})-H(Y)],
\end{equation}
where $H(Y)$ is the entropy of $Y$ induced by $P_y$.
Finally, let
\begin{equation}
\label{xis}
\xi^*(R,T)\edfn \min_{\theta,\tilde{\theta}}\max_{0\le s\le \rho\le 1}
\frac{E(\theta,\tilde{\theta},s,\rho)-\rho R-sT}
{(1-s)E_1(R,T,\theta)+sE_1(R,T,\tilde{\theta})},
\end{equation}
with the convention that if the denominator vanishes, then $\xi^*(R,T)\edfn 1$.
Our main result, in this section is the following:
\begin{theorem}
Consider the ensemble of codes where each codeword is drawn independently,
under the uniform distribution $Q(\bx)=1/|\calX|^n$ for all $\bx$. Then,
\begin{enumerate}
\item For every $\xi\le \xi^*(R,T)$, $$\limsup_{n\to\infty} \frac{1}{n}\log \bar{K}_n\le 0.$$
\item For $\xi=\xi^*(R,T)$, the average probability of $\calE_1$ and the average probability of
$\calE_2$, associated with the decoder $\hat{\calR}$, decay with exponential rates
at least as large as $\xi^*(R,T)\cdot E_1(R,T,\theta)$ 
and $\xi^*(R,T)\cdot E_1(R,T,\theta)+T$, respectively,
for all $\theta\in\Theta$.
\end{enumerate}
\end{theorem}
The proof of Theorem 2 appears in the appendix.

We now pause to discuss Theorem 2 and some of its aspects.

Theorem 2 suggests a conceptually simple strategy: Given $R$ and $T$, first compute
$\xi^*(R,T)$ using eq.\ (\ref{xis}). This may require
some non-trivial optimization procedures, but it has to be done only once,
and since this is a single--letter expression,
it can be carried at least numerically, if closed--form analytic
expressions are not apparent to be available (see the example of the BSC below).
Once $\xi^*(R,T)$ has been computed, apply the decoding rule $\hat{\calR}$ with
$\xi=\xi^*(R,T)$, and the theorem guarantees that the resulting random coding error exponents
of $\calE_1$ and $\calE_2$ are as specificed in the second item of that theorem.

The theorem is interesting, of course, only 
when $\xi^*(R,T) > 0$, which is the case
in many situations, at least as long as $R$ and $T$ are not too large. When
$\xi^*(R,T) > 0$, the proposed universal decoder 
with $\xi=\xi^*(R,T)$ has the important property that whenever Forney's optimum
decoder yields an exponential decay of 
$\overline{\mbox{Pr}}\{\calE_1\}$ ($E_1(R,T,\theta) > 0$), then so 
does the corresponding exponent of the proposed decoder, $\hat{\calR}$. It should be pointed
out that the exponential rates $\xi^*(R,T)\cdot E_1(R,T,\theta)$ 
and $\xi^*(R,T)\cdot E_1(R,T,\theta)+T$,
guaranteed by Theorem 2, are only lower bounds to the real exponential rates, and that true
exponential rate, at some points in $\Theta$, might be larger.

The derivation of $\xi^*(R,T)$ is carried out (see the appendix) using the same
bounding techniques as in Gallager's classical work and as in \cite{Forney68},
which are apparently tight in the random coding sense.
We therefore conjecture that $\xi^*(R,T)$ 
is not merely a lower bound to the best achievable
fraction of $E_1(R,T,\theta)$ that is 
universally achievable, but it actually cannot be
improved upon. If this conjecture is true, 
it means that unlike the case of ordinary
universal decoding (without erasures), 
where the optimum random coding error exponent
is universally achievable over the DMC, i.e., $\xi^*=1$
\cite{CK81},\cite{Ziv85}, here, when erasures are brought into
the picture, this is no longer the case, as $\xi^*(R,T)$ is normally striclty less than
unity, as we demonstrate later in the example of the BSC. We will also demonstrate, in
this example, that for the case $T=0$, which is asymptotically
equivalent to the case without erasures 
in the sense that $E_1(R,0,\theta)=E_2(R,0,\theta)$ coincide with Gallager's
random coding exponent \cite{Gallager68}
(although erasures are still
possible), we get $\xi^*(R,0)=1$, in agreement 
with the aforementioned full universality
result for ordinary universal decoding.

In Theorem 2, we assumed that the random coding distribution $Q$ is uniform over $\calX^n$.
This assumption is reasonable since, in the absence of any prior 
knowledge about the channel, no vectors in $\calX^n$ appear 
to have any preference
over other vectors (see also \cite{SF04} for another justification).
It is also relatively simple to analyze the random 
coding performance in this case. It is straightforward, however, to modify the
results to any random coding distribution $Q(\bx)$ which depends on $\bx$ 
only via its empirical distribution (for example, any other i.i.d.\ distribution, or a
uniform distribution within one type class). This can easily be done using the method
of types \cite{CK81} (see the appendix).

Our last comment concerns the choice of the threshold $T$.
Thus far, we assumed that $T$ is a constant, independent of $\theta$. However,
in some situations,
it makes sense to let $T$ depend on the quality of the channel, and hence on the parameter
$\theta$. Intuitively, for fixed $T$, if the 
signal--to--noise ratio (SNR) becomes very high, the erasure option will be used so
rarely, that it will effectively be non--existent. This means
that we are actually no longer ``enjoying'' the benefits of the
erasure option, and hence not the gain in the undetected error exponent that is associated with it.
An alternative 
approach is to let $T=T_\theta$ depend on $\theta$ in a certain way.
In this case, $K_n(\calC)$ would be redefined as follows:
\begin{equation}
K_n(\calC)=\min_{\calR}\max_{\theta\in\Theta}
\frac{e^{-nT_\theta}\mbox{Pr}\{\calE_1\}
+\mbox{Pr}\{\calE_2\}}{e^{-n[\xi E_1(R,T_\theta,\theta)+T_\theta]}}.
\end{equation}
The corresponding generalized version of the competitive minimax
decision rule $\hat{\calR}$, would now be:
\begin{eqnarray}
\hat{\calR}_m&=&\left\{\by:~g(\bx_m,\by)
\ge \sum_{m'\ne m} h(\bx_{m'},\by)
\right\},~m=1,\ldots,M\nonumber\\
\hat{\calR}_0&=&\bigcap_{m=1}^M \hat{\calR}_m^c,
\end{eqnarray}
where
\begin{equation}
g(\bx_m,\by)\edfn\max_\theta[P_\theta(\by|\bx_m)\cdot
e^{n\xi E_1(R,T_\theta,\theta)}]
\end{equation}
and
\begin{equation}
h(\bx_{m},\by)\edfn\max_\theta[P_\theta(\by|\bx_m)\cdot
e^{n[\xi E_1(R,T_\theta,\theta)+T_\theta]}].
\end{equation}
By extending the 
performance analysis carried 
out in the appendix, the resulting expression of $\xi^*$
now becomes
\begin{equation}
\xi^*(R)\edfn \min_{\theta,\tilde{\theta}}\max_{0\le s\le \rho\le 1}
\frac{E(\theta,\tilde{\theta},s,\rho)-\rho R-sT_{\tilde{\theta}}}
{(1-s)E_1(R,T_\theta,\theta)+sE_1(R,T_{\tilde{\theta}},\tilde{\theta})}.
\end{equation}
The main question that naturally arises, in this case, is:
which function $T_\theta$ would be reasonable to choose?
A plausible guideline could be based on the typical behavior of
$$\tau_\theta=\lim_{N\to\infty}\frac{1}{N}
\bE\ln \frac{P_\theta(\bY|\bx_m)}{\sum_{m'\ne m} P_\theta(\bY|\bx_{m'})}$$
which can be assessed, using standard bounding techniques,
under the hypothesis that $\bx_m$ is the correct message. For example, 
$T_\theta$ may be given by $\alpha\tau_\theta$ with some
constant $\alpha\in[0,1]$, or $\tau_\theta-\beta$ for some $\beta > 0$. This will make
the probability of erasure (exponentially) small, but not {\it too} small, so that there would
be some gain in the undetected error exponent for every $\theta$.

\section{Example -- the Binary Symmetric Channel}

Consider the BSC, where $\calX=\calY=\{0,1\}$, and where $\theta$ designates
the crossover probability. We would like to examine, more closely, the expression
of $\xi^*(R,T)$ and its behavior in this case. Let
$h_2(u)$ denote the binary entropy function, $-u\ln u-(1-u)\ln(1-u)$, $u\in[0,1]$.
Denoting the modulo 2 sum of $X$ and $Y$ by $X\oplus Y$, we have:
\begin{eqnarray}
F(P_y,\lambda,\theta)&=&\ln 2-\max_{P_{x|y}}[H(X|Y)+\lambda\bE\ln P(Y|X)]\nonumber\\
&=&\ln 2-\max_{P_{x|y}}\left\{H(X|Y)+
\lambda\bE\ln \left[(1-\theta)\left(\frac{\theta}{1-
\theta}\right)^{X\oplus Y}\right]\right\}\nonumber\\
&=&\ln 2-\lambda\ln(1-\theta)-\max_{P_{x|y}}\left[H(X|Y)+
(\lambda\ln\frac{\theta}{1-\theta})\cdot\bE(X\oplus Y)\right]\nonumber\\
&=&\ln 2-\lambda\ln(1-\theta)-\max_{P_{x|y}}\left[H(X\oplus Y|Y)+
(\lambda\ln\frac{\theta}{1-\theta})\cdot\bE(X\oplus Y)\right]\nonumber\\
&\ge&\ln 2-\lambda\ln(1-\theta)-\max_{P_{x|y}}\left[H(X\oplus Y)+
(\lambda\ln\frac{\theta}{1-\theta})\cdot\bE(X\oplus Y)\right]\nonumber\\
&=&\ln 2-\lambda\ln(1-\theta)-\max_u\left[h_2(u)+
(\lambda\ln\frac{\theta}{1-\theta})\cdot u\right]\nonumber\\
&=&\ln 2-\lambda\ln(1-\theta)-\ln\left[1+
\left(\frac{\theta}{1-\theta}\right)^\lambda\right]\nonumber\\
&=&\ln 2-\ln[\theta^\lambda+(1-\theta)^\lambda],
\end{eqnarray}
where the inequality is, in fact, an equality 
achieved by a backward $P_{x|y}$ where $X\oplus Y$
is independent of $Y$. Since $F(P_y,\lambda,\theta)$ is independent of $P_y$,
this easily yields 
\begin{equation}
E(\theta,\tilde{\theta},\rho,s)=\rho\ln 2-\ln[\theta^{1-s}+(1-\theta)^{1-s}]
-\rho\ln[\tilde{\theta}^{s/\rho}+(1-\tilde{\theta})^{s/\rho}]
\end{equation}
and so,
\begin{equation}
\xi^*(R,T)=\min_{\theta,\tilde{\theta}}\max_{0\le s\le \rho\le 1}\frac 
{\rho\ln 2-\ln[\theta^{1-s}+(1-\theta)^{1-s}]
-\rho\ln[\tilde{\theta}^{s/\rho}+(1-\tilde{\theta})^{s/\rho}]-\rho R -sT}
{(1-s)E_1(R,T,\theta)+sE_1(R,T,\tilde{\theta})},
\end{equation}
with
\begin{equation}
E_1(R,T,\theta)=\max_{0\le s\le\rho\le 1}
\{\rho\ln 2-\ln[\theta^{1-s}+(1-\theta)^{1-s}]
-\rho\ln[\theta^{s/\rho}+(1-\theta)^{s/\rho}]\}.
\end{equation}
This expression, although still involves
non--trivial optimizations, is much more
explicit than the general one. We next offer a few observations
regarding the function $\xi^*(R,T)$ for the example of the BSC.

First, observe that if $\Theta$ is a singleton, i.e., we are back to the
case of a known channel, then 
$\theta=\tilde{\theta}$, and the numerator, after maximization
over $\rho$ and $s$, becomes $E_1(R,T,\theta)$, 
and so does the denominator, thus
$\xi^*(R,T)=1$, as expected.
Secondly,
we argue that there exists a 
region of $R$ and $T$ (both not too large) such that
$\xi^*(R,T) > 0$. To see this, note that there are four possibilities regarding
the minimizers $\theta$ and $\tilde{\theta}$ in the above minimax problem:
\begin{enumerate}
\item $\theta=\tilde{\theta}=1/2$: In this case, the denominator vanishes too
and so, $\xi^{*}(R,T)=1$.
\item Both $\theta\ne 1/2$ and $\tilde{\theta}\ne 1/2$: Let $\hat{\theta}$ be the closer to $1/2$
between $\theta$ and $\tilde{\theta}$. Then, the numertor is obviously lower bounded by
$$\rho \ln 2-\ln[\hat{\theta}^{1-s}+(1-\hat{\theta})^{1-s}]
-\rho\ln[\hat{\theta}^{s/\rho}+(1-\hat{\theta})^{s/\rho}]-\rho R-sT,$$
which upon maximizing over $\rho$ and $s$ gives $E_1(R,T,\hat{\theta})$,
which is positive as long as $R$ and $T$ are not too large.
\item $\theta=1/2$ and $\tilde{\theta}\ne 1/2$: In this case, the numerator is given by
$$\rho \ln 2-
\rho\ln[\tilde{\theta}^{s/\rho}+(1-\tilde{\theta})^{s/\rho}]-\rho R-s(T+\ln 2).$$
Choosing $\rho=1$ and $s=1/2$, we get
$$\frac{1}{2}\ln 2-\ln\left[\sqrt{\tilde{\theta}}+\sqrt{1-\tilde{\theta}}\right]-
\left(R+\frac{T}{2}\right),$$
which is positive as long as $R$ and $T$ are not too large.
\item $\theta\ne 1/2$ and $\tilde{\theta}=1/2$: In this case, the numerator is given by
$$s\ln 2-
\ln[\theta^{1-s}+(1-\theta)^{1-s}]-\rho R-sT,$$
and once again, choosing $\rho=1$ and $s=1/2$ gives exactly the same expression as in item 3,
except that $\theta$ replaces $\tilde{\theta}$, and hence the conclusion is identical.
\end{enumerate}

We next demonstrate that $\xi^*(R,0)=1$.
Referring to the definition of the Gallager function $E(\theta,\rho)$ for the BSC:
\begin{equation}
E(\theta,\rho)=\rho\ln 2 -(1+\rho)\ln[\theta^{1/(1+\rho)}+(1-\theta)^{1/(1+\rho)}]-\rho R,
\end{equation}
let us define $\rho'=1/(1-s)-1$ and $\rho''=\rho/s-1$, and rewrite the numerator
of the expression for $\xi^*(R,0)$ as follows:
\begin{eqnarray}
&&
\rho\ln 2-\ln[\theta^{1-s}+(1-\theta)^{1-s}]
-\rho\ln[\tilde{\theta}^{s/\rho}+(1-\tilde{\theta})^{s/\rho}]-\rho R\nonumber\\
&=&
\rho\ln 2-\ln[\theta^{1/(1+\rho')}+(1-\theta)^{1/(1+\rho')}]
-\rho\ln[\tilde{\theta}^{1/(1+\rho'')}+(1-\tilde{\theta})^{1/(1+\rho'')}]-\rho R\nonumber\\
&=&\frac{1}{1+\rho'}\{
\rho'\ln 2-(1+\rho')\ln[\theta^{1/(1+\rho')}+(1-\theta)^{1/(1+\rho')}]-\rho'R\}+\nonumber\\
& &+\frac{\rho}{1+\rho''}\{\rho''\ln 2-(1+\rho'')\ln[\tilde{\theta}^{1/(1+\rho'')}+
(1-\tilde{\theta})^{1/(1+\rho'')}]-\rho''R\}\nonumber\\
&=&(1-s)E(\theta,\rho')+
sE(\tilde{\theta},\rho'')\nonumber\\
&=&(1-s)E\left(\theta,\frac{1}{1-s}-1\right)+
sE\left(\tilde{\theta},\frac{\rho}{s}-1\right).
\end{eqnarray}
Now, let us choose $s=\rho/(1+\tilde{\rho})$, where $\tilde{\rho}$ is the achiever
of $E^*(\tilde{\theta})=\max_{0\le\rho\le 1}E(\tilde{\theta},\rho)$, 
and $\rho=\rho^*(1+\tilde{\rho})/(1+\rho^*)$, where $\rho^*$
is the achiever of $E^*(\theta)=\max_{0\le\rho\le 1}E(\theta,\rho)$ 
(observing that $\rho^*(1+\tilde{\rho})/(1+\rho^*)\le 1$,
therefore this is choice is feasible). With this choice, the numerator of $\xi^*(R,T)$
becomes equal to the denominator, and so, $\xi^*(R,T)=1$.

Finally, in Table 1, we provide some numerical results pertaining to the function $\xi^*(R,T)$,
where all minimizations and maximizations were carried out by an exhaustive search with
a step-size of $0.01$ in each dimension.
As can be seen, at the left--most column, corresponding to
$T=0$, we indeed obtain $\xi^*(R,0)=1$. As can also be seen, $\xi^*(R,T)$ 
is always strictly less than unity for $T > 0$, and it in general
decreases as $T$ grows.

\begin{table}
\begin{center}
\begin{tabular}{||l|c|c|c|c|c|c|c||} \hline
$R/T$ & $T=0.000$ & $T=0.025$ & $T=0.050$ & $T=0.075$ & $T=0.100$ & $T=0.125$ & $T=0.150$\\ \hline
$R=0.00$ & 1.000 & 0.364 & 0.523 & 0.418 & 0.396 & 0.422 & 0.298 \\ \hline
$R=0.05$ & 1.000 & 0.756 & 0.713 & 0.656 & 0.535 & 0.562 & 0.495 \\ \hline
$R=0.10$ & 1.000 & 0.858 & 0.774 & 0.648 & 0.655 & 0.585 & 0.518 \\ \hline
$R=0.15$ & 1.000 & 0.877 & 0.809 & 0.720 & 0.713 & 0.662 & 0.622 \\ \hline
$R=0.20$ & 1.000 & 0.905 & 0.815 & 0.729 & 0.729 & 0.684 & 0.647 \\ \hline
$R=0.25$ & 1.000 & 0.912 & 0.832 & 0.763 & 0.706 & 0.661 & 0.627 \\ \hline
$R=0.30$ & 1.000 & 0.896 & 0.850 & 0.788 & 0.738 & 0.644 & 0.613 \\ \hline
\end{tabular}
\caption{Numerical values of $\xi^*(R,T)$ for various values of $R$ and $T$.}
\end{center}
\end{table}

\section{Conclusion}

We have addressed the problem of universal decoding with erasures, using
the competitive minimax methodology proposed in \cite{FM02}, which proved
useful. This is in contrast to earlier 
approaches for deriving universal decoders,
based on joint typicality considerations, 
for which we found no apparent extensions 
to accommodate Forney's erasure decoder.
In order to guarantee
uniform achievability of a certain fraction of the exponent, the competitive 
minimax approach was applied to the Lagrangian, pertaining to a weighted sum
of the two error probabilities.

The analysis of the minimax ratio, $\bar{K}_n$, 
resulted in a single--letter lower bound to
the largest universally achievable fraction, 
$\xi^*(R,T)$ of Forney's exponent.
We conjecture that $\xi^*(R,T)$ is a tight lower 
bound that cannot be improved upon.
In addition to the reasons we gave earlier, why we believe in this conjecture,
we have also seen that it is supported by the fact that 
at least in extreme cases, like the case $T=0$
and the case where $\Theta$ is a singleton, 
it gives the correct value $\xi^*(R,T)=1$, as expected.
An interesting problem for future work would be to prove this conjecture.
This requires the derivation of an exponentially tight lower bound to $K_n$,
which is a challenge. 

The analysis technique offerred in this paper
opens the door to similar performance analyses of 
competitive--minimax universal decoders with various types of random
coding distributions (cf.\ the second to the 
last paragraph of the discussion that
follows Theorem ~2).
This is in contrast to earlier works (see, e.g., 
\cite{CK81}, \cite{Ziv85}), which were strongly based on the assumption
that the random coding distribution is uniform within a set.
A similar analysis technique 
can be applied also to universal decoding without erasures.

Finally, we analyzed the example of the BSC 
in full detail and demonstrated that $\xi^*(R,0)=1$.
We have also provided some numerical results for this case.

\section*{Appendix -- Proof of Theorem 2}
\renewcommand{\theequation}{A.\arabic{equation}}
    \setcounter{equation}{0}

For an event $\calE\subseteq\calY^n$,
let $1\{\by|\calE\}$ denote the indicator function of $\calE$, i.e., $1\{\by|\calE\}=1$
if $\by\in\calE$ and $1\{\by|\calE\}=0$ otherwise. First, observe that
\begin{eqnarray}
1\{\by|\hat{\calR}_m\}&=&
1\left\{\by| f(\bx_m,\by)\ge e^{nT}\sum_{m'\ne m}
f(\bx_{m'},\by)\right\}\nonumber\\
&\le&\min_{0\le s\le 1}
\left[\frac{f(\bx_m,\by)}{e^{nT}\sum_{m'\ne m}
f(\bx_{m'},\by)}\right]^{1-s}
\end{eqnarray}
and similarly,
\begin{equation}
1\{\by\in\hat{\calR}_m^c\}\le
\min_{0\le s\le 1}\left[
\frac
{e^{nT}\sum_{m'\ne m}
f(\bx_{m'},\by)}
{f(\bx_m,\by)}\right]^s.
\end{equation}
Then, we have:
\begin{eqnarray}
\bar{K}_n&\le&
\bE\{K_n(\hat{\calR},\calC)\}\nonumber\\ 
&=&
\bE\left\{\max_\theta\frac{e^{-nT}\mbox{Pr}\{\calE_1\}+\mbox{Pr}\{\calE_2\}}
{e^{-n[\xi E_1(R,T,\theta)+T]}}\right\}\nonumber\\
&=&\bE\left\{\max_\theta\frac{1}{M}\sum_{m=1}^M\sum_{\by\in\calY^n}
e^{-nT}\left(P_\theta(\by|\bX_m)\cdot 
e^{n[\xi E_1(R,T,\theta)+T]}\right)\cdot 1\{\by|\hat{\calR}_m^c\}+\right. \nonumber\\
& &\left.\left(\sum_{m'\ne m}P_\theta(\by|\bX_{m'})\cdot e^{n[\xi E_1(R,T,\theta)+T]}\right)\cdot
1\{\by|\hat{\calR}_m\}\right\}\nonumber\\
&\lea&\bE\left\{\frac{1}{M}\sum_{m=1}^M\sum_{\by\in\calY^n}
e^{-nT}\max_\theta\left(P_\theta(\by|\bX_m)\cdot 
e^{n[\xi E_1(R,T,\theta)+T]}\right)\cdot 1\{\by|\hat{\calR}_m^c\}+\right. \nonumber\\
& &\left.\left(\sum_{m'\ne m}\max_\theta [P_\theta(\by|\bX_{m'})\cdot 
e^{n[\xi E_1(R,T,\theta)+T]}]\right)\cdot
1\{\by|\hat{\calR}_m\}\right\}\nonumber\\
&=&\bE\left\{\frac{1}{M}\sum_{m=1}^M\sum_{\by\in\calY^n}
e^{-nT}f(\bX_m,\by) 
\cdot 1\{\by|\hat{\calR}_m^c\}+
\left(\sum_{m'\ne m}f(\bX_{m'},\by) 
\right)\cdot
1\{\by|\hat{\calR}_m\}\right\}\nonumber\\
&\leb&\bE\left\{\frac{1}{M}\sum_{m=1}^M\sum_{\by\in\calY^n}
e^{-nT}f(\bX_m,\by) 
\cdot \min_{0\le s\le 1}\left[
\frac
{e^{nT}\sum_{m'\ne m}
f(\bX_{m'},\by)}
{f(\bX_m,\by)}\right]^s
+\right. \nonumber\\
& &\left.\left(\sum_{m'\ne m}f(\bX_{m'},\by) 
\right)\cdot
\min_{0\le s\le 1}\left[\frac{
f(\bX_m,\by)}{e^{nT}\sum_{m'\ne m}
f(\bX_{m'},\by)}\right]^{1-s}
\right\}\nonumber\\
&=&\bE\left\{\frac{2}{M}\sum_{m=1}^M\sum_{\by\in\calY^n}
\min_{0\le s\le 1}\left\{
e^{-nT(1-s)}f^{1-s}(\bX_m,\by)\left(\sum_{m'\ne m}f(\bX_{m'},\by)\right)^s
\right\}\right\}\nonumber\\
&=&\bE\left\{\frac{2}{M}\sum_{m=1}^M\sum_{\by\in\calY^n}
\min_{0\le s\le 1}\left\{
e^{-nT(1-s)}
\left(\max_{\theta\in\Theta_n} 
P_\theta(\by|\bX_m)e^{n[\xi E_1(R,T,\theta)+T]}
\right)^{1-s}\cdot\right.\right.\nonumber\\
& &\left.\left.\left(\sum_{m'\ne m}\max_{\tilde{\theta}\in\Theta_n}
P_{\tilde{\theta}}(\by|\bX_{m'})e^{n[\xi E_1(R,T,\tilde{\theta})+T]}
\right)^s\right\}\right\}\nonumber\\
&\lec&\bE\left\{\frac{2}{M}\sum_{m=1}^M\sum_{\by\in\calY^n}
\min_{0\le s\le 1}\left\{
e^{-nT(1-s)}
\left[\max_{\theta\in\Theta_n}
P_\theta(\by|\bX_m)e^{n[\xi E_1(R,T,\theta)+T]}\right]^{1-s}
\times\right.\right.\nonumber\\
& &\left.\left.\left(\sum_{m'\ne m}\sum_{\tilde{\theta}\in\Theta_n}
P_{\tilde{\theta}}(\by|\bX_{m'})e^{n[\xi E_1(R,T,\tilde{\theta})+T]}
\right)^s\right\}\right\}\nonumber\\
&=&\bE\left\{\frac{2}{M}\sum_{m=1}^M\sum_{\by\in\calY^n}
\min_{0\le s\le 1}\left\{
e^{-nT(1-s)}
\left[\max_{\theta\in\Theta_n}
P_\theta(\by|\bX_m)e^{n[\xi E_1(R,T,\theta)+T]}\right]^{1-s}
\times\right.\right.\nonumber\\
& &\left.\left.\left(\sum_{\tilde{\theta}\in\Theta_n}\sum_{m'\ne m}
P_{\tilde{\theta}}(\by|\bX_{m'})e^{n[\xi E_1(R,T,\tilde{\theta})+T]}
\right)^s\right\}\right\}\nonumber\\
&\led&\bE\left\{\frac{2}{M}\sum_{m=1}^M\sum_{\by\in\calY^n}
\min_{0\le s\le 1}\left\{
e^{-nT(1-s)}
\left[\max_{\theta\in\Theta_n}
P_\theta(\by|\bX_m)e^{n[\xi E_1(R,T,\theta)+T]}\right]^{1-s}
\times\right.\right.\nonumber\\
& &\left.\left.\left(|\Theta_n|\cdot\max_{\tilde{\theta}\in\Theta_n}\sum_{m'\ne m}
P_{\tilde{\theta}}(\by|\bX_{m'})e^{n[\xi E_1(R,T,\tilde{\theta})+T]}
\right)^s\right\}\right\}\nonumber\\
&\le&\bE\left\{\frac{2|\Theta_n|}{M}\sum_{m=1}^M\sum_{\by\in\calY^n}
\min_{0\le s\le 1}\left\{
e^{-nT(1-s)}
\left[\max_{\theta\in\Theta_n}
P_\theta(\by|\bX_m)e^{n[\xi E_1(R,T,\theta)+T]}\right]^{1-s}
\times\right.\right.\nonumber\\
& &\left.\left.\left(\max_{\theta\in\Theta_n}\sum_{m'\ne m}
P_{\tilde{\theta}}(\by|\bX_{m'})e^{n[\xi E_1(R,T,\tilde{\theta})+T]}
\right)^s\right\}\right\}\nonumber\\
&\lee&\bE\left\{\frac{2|\Theta_n|}{M}\sum_{m=1}^M\sum_{\by\in\calY^n}
\sum_{\theta\in\Theta_n}\sum_{\tilde{\theta}\in\Theta_n}
\min_{0\le s\le 1}\left\{
e^{-nT(1-s)}\left[
P_\theta(\by|\bX_m)e^{n[\xi E_1(R,T,\theta)+T]}\right]^{1-s}
\times\right.\right.\nonumber\\
& &\left.\left.\left(\sum_{m'\ne m}
P_{\tilde{\theta}}(\by|\bX_{m'})e^{n[\xi E_1(R,T,\tilde{\theta})+T]}
\right)^s\right\}\right\}\nonumber\\
&=&\frac{2|\Theta_n|}{M}
\sum_{\theta\in\Theta_n}\sum_{\tilde{\theta}\in\Theta_n}
\sum_{m=1}^M\sum_{\by\in\calY^n}
\bE\min_{0\le s\le 1}\left\{
e^{-nT(1-s)}\left[
P_\theta(\by|\bX_m)e^{n[\xi E_1(R,T,\theta)+T]}\right]^{1-s}
\times\right.\nonumber\\
& &\left.\left(\sum_{m'\ne m}
P_{\tilde{\theta}}(\by|\bX_{m'})e^{n[\xi E_1(R,T,\tilde{\theta})+T]}
\right)^s\right\}\nonumber\\
&\lef&\frac{2|\Theta_n|^3}{M}
\max_{\theta\in\Theta_n}\max_{\tilde{\theta}\in\Theta_n}
\sum_{m=1}^M\sum_{\by\in\calY^n}
\bE\min_{0\le s\le 1}\left\{
e^{-nT(1-s)}\left[
P_\theta(\by|\bX_m)e^{n[\xi E_1(R,T,\theta)+T]}\right]^{1-s}\times
\right.\nonumber\\
& &\left.\left(\sum_{m'\ne m}
P_{\tilde{\theta}}(\by|\bX_{m'})e^{n[\xi E_1(R,T,\tilde{\theta})+T]}
\right)^s\right\}\nonumber\\
&\le&\frac{2|\Theta_n|^3}{M}
\max_{\theta\in\Theta}\max_{\tilde{\theta}\in\Theta}
\min_{0\le s\le 1}
\sum_{m=1}^M\sum_{\by\in\calY^n}
\bE\left\{
e^{-nT(1-s)}\left[
P_\theta(\by|\bX_m)e^{n[\xi E_1(R,T,\theta)+T]}\right]^{1-s}
\times\right.\nonumber\\
& &\left.\left(\sum_{m'\ne m}
P_{\tilde{\theta}}(\by|\bX_{m'})e^{n[\xi E_1(R,T,\tilde{\theta})+T]}
\right)^s\right\}\nonumber\\
&=&\frac{2|\Theta_n|^3}{M}
\max_{\theta\in\Theta}\max_{\tilde{\theta}\in\Theta}
\min_{0\le s\le 1} e^{n[\xi\{(1-s)E_1(R,T,\theta)+sE_1(R,T,\tilde{\theta})\}+sT]}\times\nonumber\\
& &\sum_{m=1}^M\sum_{\by\in\calY^n}
\bE\left\{
P_\theta^{1-s}(\by|\bX_m)
\cdot\left(\sum_{m'\ne m}
P_{\tilde{\theta}}(\by|\bX_{m'})
\right)^s\right\},
\end{eqnarray}
where (a) follows from the fact that the maximum (over $\theta$)
of a summation is upper bounded by the summation of the maxima,
(b) follows from (A.1) and (A.2), and (c), (d), (e) and (f) all follow
from the fact that if $g(\theta)$ is non--negative then
$$\max_{\theta\in\Theta_n}g(\theta)\le\sum_{\theta\in\Theta_n}g(\theta)
\le|\Theta_n|\cdot\max_{\theta\in\Theta_n}g(\theta).$$
Assuming that the codewords are drawn independently, we then have:
\begin{eqnarray}
\bar{K}_n&\le&
\frac{2|\Theta_n|^3}{M}
\max_{\theta\in\Theta}\max_{\tilde{\theta}\in\Theta}
\min_{0\le s\le 1} e^{n[\xi\{(1-s)E_1(R,T,\theta)+sE_1(R,T,\tilde{\theta})\}+sT]}\times\nonumber\\
& &\sum_{m=1}^M\sum_{\by\in\calY^n}
\bE\{
P_\theta^{1-s}(\by|\bX_m)\}
\cdot\bE\left\{\left(\sum_{m'\ne m}
P_{\tilde{\theta}}(\by|\bX_{m'})
\right)^s\right\}\nonumber\\
&=&\frac{2|\Theta_n|^3}{M}
\max_{\theta\in\Theta}\max_{\tilde{\theta}\in\Theta}
\min_{0\le s\le 1} e^{n[\xi\{(1-s)E_1(R,T,\theta)+sE_1(R,T,\tilde{\theta})\}+sT]}\times\nonumber\\
& &\sum_{m=1}^M\sum_{\by\in\calY^n}\bE\{
P_\theta^{1-s}(\by|\bX_m)\}
\cdot\min_{s\le\rho\le 1}\bE\left\{\left(\left[\sum_{m'\ne m}
P_{\tilde{\theta}}(\by|\bX_{m'})\right]^{s/\rho}
\right)^\rho\right\}\nonumber\\
&\le&\frac{2|\Theta_n|^3}{M}
\max_{\theta\in\Theta}\max_{\tilde{\theta}\in\Theta}
\min_{0\le s\le 1} e^{n[\xi\{(1-s)E_1(R,T,\theta)+sE_1(R,T,\tilde{\theta})\}+sT]}\times\nonumber\\
& &\sum_{m=1}^M\sum_{\by\in\calY^n}\bE\{
P_\theta^{1-s}(\by|\bX_m)\}
\cdot\min_{0\le s\le\rho\le 1}\bE\left[\left(\sum_{m'\ne m}
P_{\tilde{\theta}}^{s/\rho}(\by|\bX_{m'})
\right)^\rho\right\}\nonumber\\
&\le&\frac{2|\Theta_n|^3}{M}
\max_{\theta\in\Theta}\max_{\tilde{\theta}\in\Theta}
\min_{0\le s\le \rho\le 1} 
e^{n[\xi\{(1-s)E_1(R,T,\theta)+sE_1(R,T,\tilde{\theta})\}+sT]}\times\nonumber\\
& &\sum_{m=1}^M\sum_{\by\in\calY^n}\bE\{
P_\theta^{1-s}(\by|\bX_m)\}
\cdot\left(\sum_{m'\ne m}\bE\{
P_{\tilde{\theta}}^{s/\rho}(\by|\bX_{m'})\}
\right)^\rho,
\end{eqnarray}
where in the last step we have used Jensen's inequality.
Now, observe that the summands do not depend on $m$, therefore, the 
effects of the summation over $m$ and
the factor of $1/M$ cancel each other. Also, the sum of $M-1$ contributions
of identical expectations $\bE\{P_{\tilde{\theta}}^{s/\rho}(\by|\bX_{m'})\}$
creat a factor of $M-1$ (upper bounded by $M$) raised to the power of $\rho$.
Denoting 
$$U(\by,\lambda,\theta)=\bE\{P_\theta^\lambda(\by|\bX)\},$$
we have:
\begin{eqnarray}
\bar{K}_n&\le&
2|\Theta_n|^3
\max_{\theta\in\Theta}\max_{\tilde{\theta}\in\Theta}
\min_{0\le s\le \rho\le 1} M^\rho\cdot
e^{n[\xi\{(1-s)E_1(R,T,\theta)+sE_1(R,T,\tilde{\theta})\}+sT]}\times\nonumber\\
& &\sum_{\by\in\calY^n}U(\by,1-s,\theta)\cdot U^\rho(\by,s/\rho,\tilde{\theta}).
\end{eqnarray}
To compute $U(\by,\lambda,\theta)$, 
we use the method of types \cite{CK81}. 
Now, $Q$ is assumed i.i.d.\ and uniform over the
entire input space, i.e., $Q(\bx)=1/|\calX|^n$ for all $\bx$. Let $\hat{P}_{\bx\by}$
denote the empirical joint distribution of $(\bx,\by)$ and let $\hat{\bE}_{\bx\by}\{\cdot\}$
denote the corresponding empirical expectation, i.e., the expectation w.r.t.\ $\hat{P}_{\bx\by}$.
Also, let $T(\bx|\by)$ denote the conditional type class of $\bx$ given $\by$, i.e.,
the set of $\bx'$ with $\hat{P}_{\bx'\by}=\hat{P}_{\bx\by}$ and let
$\hat{H}_{\bx\by}(X|Y)$ denote the corresponding empirical conditional entropy of $X$ given $Y$.
Then,
\begin{eqnarray}
U(\by,\lambda,\theta)&=&\frac{1}{|\calX|^n}\sum_{\bx\in\calX^n}P_\theta^\lambda(\by|\bx)\nonumber\\
&=&\frac{1}{|\calX|^n}\sum_{T(\bx|\by)\subset\calX^n}
|T(\bx|\by)|\cdot e^{\lambda n\hat{\bE}_{\bx\by}\ln P_\theta(Y|X)}\nonumber\\
&\le&\frac{1}{|\calX|^n}\sum_{T(\bx|\by)\subset\calX^n}
e^{n\hat{H}_{\bx\by}(X|Y)}\cdot e^{\lambda n\hat{\bE}_{\bx\by}\ln P_\theta(Y|X)}\nonumber\\
&\le& (n+1)^{|\calY|\cdot(|\calX|-1)}\cdot e^{-nF(\hat{P}_{\by},\lambda,\theta)},
\end{eqnarray}
where
$F(P_y,\lambda,\theta)$ is defined as in eq.\ (\ref{F}).
On substituting this bound into the upper bound on $K_n(\hat{\calR})$, we get:
\begin{eqnarray}
\bar{K}_n&\le&
2|\Theta_n|^3(n+1)^{2|\calY|\cdot(|\calX|-1)}
\max_{\theta\in\Theta}\max_{\tilde{\theta}\in\Theta}
\min_{0\le s\le \rho\le 1}\nonumber\\
& &M^\rho\cdot
e^{n[\xi\{(1-s)E_1(R,T,\theta)+sE_1(R,T,\tilde{\theta})\}+sT]}
\cdot\sum_{\by\in\calY^n} e^{-n[F(P_{\by},1-s,\theta)+
\rho F(P_{\by},s/\rho,\tilde{\theta})]}\nonumber\\
&\le&2|\Theta_n|^3(n+1)^{2|\calY|\cdot(|\calX|-1)}
\max_{\theta\in\Theta}\max_{\tilde{\theta}\in\Theta}
\min_{0\le s\le \rho\le 1}\nonumber\\ 
& &M^\rho\cdot
e^{n[\xi\{(1-s)E_1(R,T,\theta)+sE_1(R,T,\tilde{\theta})\}+sT]}\cdot
\sum_{T_{\by}\subset\calY^n} e^{n\hat{H}_{\by}(Y)}\cdot e^{-n[F(P_{\by},1-s,\theta)+
\rho F(P_{\by},s/\rho,\tilde{\theta})]}\nonumber\\
&\le&2|\Theta_n|^3(n+1)^{3|\calY|\cdot(|\calX|-1)}
\max_{\theta\in\Theta}\max_{\tilde{\theta}\in\Theta}
\min_{0\le s\le \rho\le 1}\nonumber\\
& &M^\rho\cdot
e^{n[\xi\{(1-s)E_1(R,T,\theta)+sE_1(R,T,\tilde{\theta})\}+sT]}\cdot
e^{-n\min_{P_y}[F(P_y,1-s,\theta)+
\rho F(P_y,s/\rho,\tilde{\theta})-H(Y)]}\nonumber\\
&\le&2|\Theta_n|^3(n+1)^{3|\calY|\cdot(|\calX|-1)}
\max_{\theta\in\Theta}\max_{\tilde{\theta}\in\Theta}
\min_{0\le s\le \rho\le 1}\nonumber\\ 
& &M^\rho\cdot
e^{n[\xi\{(1-s)E_1(R,T,\theta)+sE_1(R,T,\tilde{\theta})\}+sT]}\cdot
e^{-nE(\theta,\tilde{\theta},s,\rho)}.
\end{eqnarray}
We would like to find the maximum value of $\xi$
such that $\bar{K}_n$ would be guaranteed not to grow exponentially.
To this end, we can now ignore the factor $2|\Theta_n|^3(n+1)^{3|\calY|\cdot(|\calX|-1)}$,
which is polynomial in $n$ (cf.\ eq.\ (\ref{Thetan})). 
Thus, the latter upper bound will be sub--exponential in $n$
as long as
\begin{equation}
\min_{\theta,\tilde{\theta}}\max_{0\le s\le \rho\le 1}[E(\theta,\tilde{\theta},s,\rho)
-\xi\{(1-s)E_1(R,T,\theta)+sE_1(R,T,\tilde{\theta})\}-\rho R-sT]\ge 0,
\end{equation}
or, equivalently, for every $(\theta,\tilde{\theta})$, there exist $(\rho,s)$,
$0\le s\le \rho\le 1$, such that
\begin{equation}
E(\theta,\tilde{\theta},s,\rho)\ge
\xi\{(1-s)E_1(R,T,\theta)+sE_1(R,T,\tilde{\theta})\}+\rho R+sT,
\end{equation}
i.e.,
\begin{equation}
\xi\le \frac{E(\theta,\tilde{\theta},s,\rho)-\rho R-sT}
{(1-s)E_1(R,T,\theta)+sE_1(R,T,\tilde{\theta})}.
\end{equation}
In other words, for every $\xi\le \xi^*(R,T)$, where $\xi^*(R,T)$ 
is defined as in eq.\ (\ref{xis})),
$K_n(\hat{\calR})$ is guaranteed 
not to grow exponentially with $n$. This completes the proof of
Theorem 2.

\end{document}